\documentclass[pra, twocolumn,showpacs,nofootinbibfloatfix,amsmath,amsfonts,amssymb]{revtex4-1}%
\usepackage{amsmath,amsfonts,amssymb,color}
\usepackage{amsthm}
\usepackage{leftidx}
\usepackage{graphicx}
\usepackage{xcolor}
\usepackage{dcolumn}
\usepackage{bm}
\usepackage{epstopdf}
\usepackage{epsfig}
\usepackage{mathdots} 
\begin{document}
	\title{Floquet Bosonic Kitaev Chain}
	\author{Raditya Weda Bomantara}
	\email{Raditya.Bomantara@kfupm.edu.sa}
	\affiliation{%
		Department of Physics, Interdisciplinary Research Center for Advanced Quantum Computing, King Fahd University of Petroleum and Minerals, 31261 Dhahran, Saudi Arabia
	}
	\date{\today}
	
	%%%%%%%%%%%%%%%%%%%% ABSTRACT %%%%%%%%%%%%%%%%%%%%%%%%
	%\begin{linenumbers}
	
	\vspace{2cm}
	
\begin{abstract}
We propose a class of periodically driven (Hermitian) modified bosonic Kitaev chains that effectively hosts rich nonHermitian Floquet topological phenomena. Two particular models are investigated in details as case studies. The first of these represents a minimal topologically nontrivial model in which nonHermitian skin effect, topological zero modes, and topological $\pi$ modes coexist. The other displays a more sophisticated model that supports multiple topological zero modes and topological $\pi$ modes in a tunable manner. By subjecting both models to perturbations such as a finite onsite bosonic frequency and spatial disorder, these features exhibit distinct responses. In particular, while generally all topological edge modes are robust against such perturbations, the nonHermitian skin effect is easily suppressed and revived by, respectively, the onsite bosonic frequency and spatial disorder in the first model, but it could be insensitive to both perturbations in the second model. Our studies thus demonstrate the prospect of a periodically driven bosonic Kitaev chain as a starting point in exploring various nonHermitian Floquet topological phases through the lens of a Hermitian system.      
	
\end{abstract}

\maketitle

\section{Introduction} 
\label{intro}

The Kitaev chain \cite{Kit} is a powerful toy model that demonstrates the possibility of Majorana fermions, the otherwise hypothetical quasiparticles, to manifest in a physical system. Despite its mathematical simplicity, its main ingredient, i.e., the $p$-wave superconductivity, is not easily achievable in experiments. Consequently, follow-up theoretical proposals \cite{Lutchyn10,Oreg10} and extensive experimental efforts \cite{Mourik12,Rokhinson12,Deng12,Finck13,Albrecht16,Chen17,Deng16,Suominen17,Nichele17,Laroche19,Liu20,Manna20,Chen20} have been made since the last two decades to realize a model that at least simulates the Kitaev chain to some degree, but it importantly supports the predicted Majorana fermions in the form of zero energy excitations termed Majorana zero modes (MZMs). The interest in MZMs is further amplified by their expected ability to encode qubits nonlocally and support topologically protected quantum gate operations \cite{Nayak08,Ahlbrecht09,Pachos12,Lahtinen17}. For this reason, Majorana-based quantum chips \cite{Karzig17,Plugge17} and quantum error correction schemes \cite{Landau16,Plugge16,Vijay17,Bomantara20a} have been theoretically proposed in recent years. On the other hand, experimental progress is presently at an impasse due to the difficulty in detecting MZMs unambiguously \cite{Pan20,Sarma21}.

Recently, the bosonic version of the Kitaev chain was proposed and extensively studied in Ref.~\cite{McDonald18}. Such a bosonic Kitaev chain (BKC) has since garnered the interest of theoreticians and experimentalists alike, resulting in a number of follow-up studies \cite{Ugh24,Slim24,Busnaina24,He25}. Indeed, unlike its fermionic counterpart, realizing the ``bosonic" version of $p$-wave pairing term is not as elusive, which is thus attractive from an experimentallist's point of view. Moreover, despite being a Hermitian many-body model, the dynamics of the BKC could be effectively described by a nonHermitian matrix \cite{McDonald18}. This in turn leads to the emergence of phenomena typically associated with a nonHermitian system. This includes the so called nonHermitian skin effect (NHSE), which is marked by the localization of all eigenstates to a system's end \cite{Yao18,Kunst18,Yao18b,Lee16,Alvarez18,Xiong18,Lin23}. The BKC thus serves as a promising platform for exploring nonHermitian physics, which remains a timely field of studies \cite{Rudner09,Hu11,Esaki11,Schomerus13,Shen18,Gliozzi24,Okuma23,Shen24,Nakamura25,Hamanaka24,Gallo25}, through the lens of a Hermitian system.  

Despite the mathematical similarity between the two types of Kitaev chain, their topology is significantly distinct. In particular, while a fermionic Kitaev chain is a topological model that could support a robust pair of MZMs, the BKC lacks the necessary feature to yield similar topologically protected edge modes. The latter is attributed to the fact that bosonic particles are mutually commuting rather than mutually anticommuting like fermionic particles are. For this reason, a recent work \cite{Bomantara25} presents a variation of the BKC that imbues it with the necessary topology to enable the formation of topological edge modes. Such a modified BKC is then shown to exhibit an effective nonHermitian topological phase in which NHSE and zero energy excitation edge modes coexist.

This work takes a step further in modifying the BKC by incorporating periodic driving. Indeed, periodic driving has a long history of enriching a system's topology, often with no static counterpart \cite{Rudner13,Nathan15}. Consequently, the studies of such Floquet topological phases \cite{Oka09,Kitagawa10,Lindner11,Ho12,Cayssol13,Ho14,Zhou14,Kundu13,Bomantara16,Bomantara16b,Fulga16,Rieder18,Jotzu14,Aidelsburger15,Flaschner16,Kitagawa12,Rechtsman13,Hu15,Zhou18,Zhou18b,Cheng19,McIver20,Tan20,Yang21,Ghuneim25} have since remained an active research field. Of particular interest and relevance to fermionic Kitaev chain is the presence of the so-called Majorana $\pi$ modes (MPMs) \cite{Jiang11,Liu13,Wang17,Bomantara20,Bomantara21,Bomantara22}, which complement MZMs in quantum computations \cite{Bomantara18,Bauer19,Bomantara18b,Bomantara20b,Matthies22}, thereby potentially reducing space overhead. More remarkably, appropriately designed periodic driving even allows for multiple MZMs and MPMs to emerge on one end of the system \cite{Zhou20,Bomantara20c}, a feature which is especially advantageous for implementing quantum error corrections \cite{Bomantara20c}. While a periodically driven fermionic Kitaev chain and its resulting MZMs and MPMs have been extensively studied, to the best of our knowledge, a periodically driven BKC is still largely unexplored and elusive. On the other hand, the interplay between the effective nonHermiticy of the BKC and periodic driving is expected to yield various Floquet nonHermitian topological phases, which are generally even richer than their Hermitian counterparts \cite{Zhou18c,Zhou19,Zhou19b,Zhang20,Wu20,Cao21,Wu21,Vyas21,Chowdhury21,Weidemann22,Zhou22,Zhou23}.

Motivated by the aforementioned advances in Floquet topological phases and the effective nonHermitian nature of the BKC, this work presents a mathematically simple class of time-periodic variations of the modified BKC introduced in Ref.~\cite{Bomantara25}. Specifically, two concrete models from this class will be considered and extensively studied to uncover different phenomena that result from the interplay among effective nonHermiticy, topology, and periodic driving. The first model represents a minimal periodically driven modified BKC that hosts topological zero and $\pi$ edge modes, i.e., the bosonic analogues of MZMs and MPMs respectively, whereas the second model involves a slightly more elaborate driving scheme that allows such zero and $\pi$ modes to proliferate in a controllable manner on one end of the system. In addition to supporting both zero and $\pi$ edge modes, we find that both models also exhibit NHSE that respond differently to the presence of finite onsite bosonic frequency and spatial disorder. Specifically, NHSE in the first model is sensitive to the presence of both onsite bosonic frequency and spatial disorder, such that the former breaks NHSE, whilst the latter gradually recovers it. By contrast, NHSE in the second model is insensitive to both effects at general parameter values. 

This paper is structured as follows. In Sec.~\ref{static}, we review the formation of effective nonHermiticity in an otherwise Hermitian BKC. Section~\ref{TP} generalizes the theory to the time-periodic realm by utilizing machinery from Floquet theory. There, important notations and definitions that are used throughout the remainder of the paper are also introduced. Section~\ref{FMBKC} presents a class of periodically driven modified BKC that will be the main subject of this paper. Two case studies from such a class are examined in detail in Secs.~\ref{Mod1} and \ref{Mod2}. In Sec.~\ref{Mod1}, a minimal model that displays NHSE while simultaneously supporting topological zero and $\pi$ edge modes, termed model 1, is introduced. In Sec.~\ref{Mod2}, we consider a different model, termed model 2, that also displays NHSE, but it is simultaneously capable of hosting multiple topological zero and $\pi$ modes in a tunable manner. In both Sec.~\ref{Mod1} and Sec.~\ref{Mod2}, the NHSE is confirmed by explicitly contrasting quasienergy excitation spectra of the models under open boundary conditions (OBC) and periodic boundary conditions (PBC), as well as by directly verifying the localization of all quasienergy excitation eigenstates on one end of the system. Meanwhile, the topological nature of the zero and $\pi$ edge modes is established by appropriately defining and calculating relevant topological invariants. In Sec.~\ref{onsite}, the effect of a nonzero onsite bosonic frequency on each model is explored. The effect of spatial disorder is then extensively discussed in Sec.~\ref{disorder}. Finally, Sec.~\ref{conc} summarizes the findings of this paper and discusses prospects for future studies.

\section{Effective nonHermitian description of Hermitian bosonic lattices}

\subsection{Time-independent case}
\label{static}

Consider a one-dimensional (1D) lattice of $N$ lattice sites, and let $a_j$ be the bosonic operator acting on site $j$. Suppose then that the system Hamiltonian contains the term
\begin{equation}
    H = H_0 + \sum_{j=1}^{N-1} \left(\Delta a_j^\dagger a_{j+1}^\dagger +h.c. \right) .
\end{equation}
Note that $H$ is clearly Hermitian (assuming $H_0$ is Hermitian). In terms of the conjugate variables $x_j$ and $p_j$ that satisfy $a_j =\frac{x_j+ \mathrm{i} p_j}{\sqrt{2}}$, it reads
\begin{eqnarray}
    H &=& H_0 + \sum_{j=1}^{N-1} \Delta_i \left( x_j p_{j+1} +p_j x_{j+1} \right) \nonumber \\
    && +\sum_{j=1}^{N-1} \Delta_r \left( x_j x_{j+1} - p_j p_{j+1} \right) ,
\end{eqnarray}
where $\Delta=\Delta_r +\mathrm{i} \Delta_i$. It is easily verified that 
\begin{eqnarray}
    [H,x_1] &=& [H_0,x_1] -\mathrm{i}\; \Delta_i x_2 +\mathrm{i}\; \Delta_r p_2 , \nonumber \\
    \left[H,p_1\right] &=& [H_0,p_1] +\mathrm{i} \; \Delta_i p_2 +\mathrm{i} \; \Delta_r x_2  , \nonumber \\
    \left[H,x_{j>1}\right] &=& [H_0,x_{j>1}] -\mathrm{i} \;\Delta_i (x_{j-1} + x_{j+1})  \nonumber \\
    && +\mathrm{i} \; \Delta_r (p_{j-1} + p_{j+1}) , \nonumber \\
    \left[H,p_{j>1}\right] &=& [H_0,p_{j>1}] +\mathrm{i} \; \Delta_i (p_{j+1} + p_{j-1}) \nonumber \\
    && +\mathrm{i} \; \Delta_r (x_{j-1} + x_{j+1}) . \label{crel}
\end{eqnarray}
Let $q_\epsilon=c_1 x_1 + \cdots +c_N x_N+d_1 p_1 +\cdots + d_N p_N$ satisfy the ``eigenvalue" equation
\begin{equation}
    [H,q_\epsilon] = \epsilon q_\epsilon \label{excivalue}
\end{equation}
for some c-numbers $c_1,\cdots,c_N,d_1,\cdots,d_N$, and $\epsilon$. Physically, $q_\epsilon$ represents an $\epsilon$-energy excitation mode, which maps any energy $E$ eigenstate $|E\rangle$ of $H$ into another eigenstate $q|E\rangle \propto |E+\epsilon\rangle$ of $H$ corresponding to energy $E+\epsilon$.

Using Eq.~(\ref{crel}), we may turn Eq.~(\ref{excivalue}) into the matrix eigenvalue equation
\begin{eqnarray}
    \mathcal{H} |q_\epsilon\rangle  &=& \epsilon |q_\epsilon\rangle ,
\end{eqnarray}
where 
\begin{eqnarray}
    |q_\epsilon\rangle &=& \left(c_1,\cdots ,c_N,d_1, \cdots ,d_N\right)^T , \nonumber \\
    \mathcal{H} &=& \mathcal{H}_0-\sum_{j=1}^{N-1} \mathrm{i} \left( \Delta_i \sigma_z  - \Delta_r \sigma_x \right) \left(|j\rangle \langle j+1 | +h.c.\right) , \nonumber \\ \label{effstate}
\end{eqnarray}
$\sigma$'s are the Pauli matrices connecting the $c$- and $d$-coefficients. Therefore, the full energy excitation spectrum of the system could then be obtained by simply diagonalizing $\mathcal{H}$. In the remainder of this paper, we shall refer to $\mathcal{H}$ as the \emph{excitation Hamiltonian} of the system.

Interestingly, even though the system Hamiltonian $H$ is Hermitian, $\mathcal{H}$ is clearly a nonHermitian matrix. Here, the nonHermiticity arises due to the number nonconserving term $\propto a_j^\dagger a_{j+1}^\dagger$ in the system Hamiltonian, which corresponds to parametric driving \cite{McDonald18}. In fact, it is easily verified that number conserving terms such as $a_j^\dagger a_j$ or $a_j^\dagger a_{j+1}$ lead to Hermitian terms in $\mathcal{H}$. By contrast, another number nonconserving term of the form $\left(a_j^\dagger\right)^2 + \left(a_j\right)^2$ yields nonHermitian terms in the excitation Hamiltonian matrix of the form $i\sigma_x |j\rangle \langle j|$ and $i\sigma_z |j \rangle \langle j |$. Generally, a quadratic number nonconserving term in the system Hamiltonian generates nonHermitian terms involving $\sigma_x \cos(n \hat{k})$ and $\sigma_z \cos(n \hat{k})$ ($n$ being an integer and $\hat{k}$ being the quasimomentum operator such that $e^{\pm \mathrm{i} \hat{k}}$ shifts the lattice site by one unit) in the corresponding excitation Hamiltonian; it is impossible to generate a nonHermitian term $\propto \sigma_y$ or any term $\propto \sin(n\hat{k})$ in the excitation Hamiltonian under a Hermitian bosonic lattice system. Such a constraint results as a consequence of the commutation relations $[x_i,p_j]=\mathrm{i} \delta_{i,j}$ and $[x_i,x_j]=[p_i,p_j]=0$.

\subsection{Time periodic case}
\label{TP}

Consider a time periodic Hamiltonian that satisfies $H(t+T)=H(t)$ for some period $T$. According to Floquet theory \cite{Shirley65,Sambe73}, many of the system's physical properties are encoded by its one-period time evolution operator (also referred to as its Floquet operator), i.e.,
\begin{equation}
    U_T=\mathcal{T} e^{-\mathrm{i} \int_0^T H(t) dt} ,
\end{equation}
where $\mathcal{T}$ is the time ordering operator. If $H(t)$ is Hermitian, $U_T$ is a unitary operator and its eigenvalues take the form of $e^{-\mathrm{i} \varepsilon T}$ for some real $\varepsilon$. In this case, the so-called quasienergy $\varepsilon$ is the time-periodic analogue of energy that is only uniquely defined within the Floquet Brillouin zone $[-\pi/T, \pi/T)$, i.e., $\varepsilon +2\pi/T$ and $\varepsilon$ describe the same physics. The corresponding eigenstate $|\varepsilon \rangle$ is then referred to as the quasienergy eigenstate.

The time periodic analogue of Eq.~(\ref{excivalue}) is the relation
\begin{equation}
    U_T q_\epsilon U_T^\dagger = e^{-\mathrm{i}\epsilon T} q_\epsilon . \label{Fexcivalue}
\end{equation}
Indeed, it is easily verified that $q_\epsilon$ maps any quasienergy eigenstate $|\varepsilon \rangle$ to another quasienergy eigenstate $q_\epsilon |\varepsilon \rangle \propto |\varepsilon +\epsilon \rangle$. Therefore, $q_\epsilon$ that satisfies Eq.~(\ref{Fexcivalue}) could be referred to as the $\epsilon$-quasienergy excitation mode. 

For a class of time-periodic Hamiltonians that can be written as
\begin{equation}
    H(t) = \begin{cases}
        H_1 & \text{ for } t\;{\rm mod} \;T<T/2 \\
        H_2 & \text{ for } t\;{\rm mod} \;T>T/2
    \end{cases} ,
\end{equation}
where $H_1$ and $H_2$ are time-independent, the Floquet operator takes a particularly simple form
\begin{equation}
    U_T = e^{-\mathrm{i} H_2 T/2} e^{-\mathrm{i} H_1 T/2}  .
\end{equation}
Equation~(\ref{Fexcivalue}) then becomes
\begin{equation}
    e^{-\mathrm{i} H_2 T/2} e^{-\mathrm{i} H_1 T/2} U_T q_\epsilon e^{\mathrm{i} H_1 T/2} e^{\mathrm{i} H_2 T/2} = e^{-\mathrm{i}\epsilon} q_\epsilon . 
\end{equation}
Next, we note the well-known identity
\begin{equation}
    e^{-\mathrm{i} \theta H} A e^{\mathrm{i} \theta H} = e^{-\mathrm{i} \theta \mathcal{C}_H} A , \label{idconj}
\end{equation}
where $\mathcal{C}_H=[H,\cdots]$ is the commutator superoperator. Specifically,
\begin{equation}
    e^{-\mathrm{i} \theta \mathcal{C}_H} A = A -\mathrm{i} \theta [H,A] - \theta^2 \frac{[H,[H,A]]}{2!} +\cdots .
\end{equation}
Equation~(\ref{Fexcivalue}) could then be equivalently written as
\begin{equation}
    u_T q_\epsilon = e^{-\mathrm{i} \epsilon T} q_\epsilon , \label{Fexcivalue2}
\end{equation}
where
\begin{equation}
    u_T = e^{-\mathrm{i} \frac{T}{2} \mathcal{C}_{H_2}} e^{-\mathrm{i} \frac{T}{2} \mathcal{C}_{H_1}} \label{FloSO}
\end{equation}
shall be referred to as \emph{the Floquet superoperator}. 

Suppose $H_1$ and $H_2$ describe a 1D bosonic lattice that take the form
\begin{eqnarray}
    H_1 &=& H_{0,1} + \left(\Delta_1 a_j^\dagger a_{j+1}^\dagger +h.c. \right) , \nonumber \\
    H_2 &=& H_{0,2} + \left(\Delta_2 a_j^\dagger a_{j+1}^\dagger +h.c. \right) . 
\end{eqnarray}
By further writing $q_\epsilon=c_1 x_1 +\cdots + c_N x_N +d_1 p_1 + \cdots + d_N p_N$, $\Delta_1 = \Delta_{1,r}+\mathrm{i} \Delta_{1,i}$, $\Delta_2 = \Delta_{2,r}+\mathrm{i} \Delta_{2,i}$, and using Eq.~(\ref{crel}), Eq.~(\ref{Fexcivalue2}) can be written as a matrix eigenvalue equation
\begin{equation}
    \mathcal{U}_T |q_\epsilon \rangle= e^{-\mathrm{i} \epsilon T} |q_\epsilon\rangle , \label{mrep}
\end{equation}
where 
\begin{equation}
    \mathcal{U}_T = e^{-\mathrm{i} \frac{T}{2} \mathcal{H}_2} e^{-\mathrm{i} \frac{T}{2} \mathcal{H}_1} , \label{FloSOm}
\end{equation}
$|q_\epsilon\rangle$ is as given in Eq.~(\ref{effstate}) and
\begin{eqnarray}
    \mathcal{H}_1 &=& \mathcal{H}_{0,1}-\sum_{j=1}^{N-1} \mathrm{i} \left( \Delta_{1,i} \sigma_z  - \Delta_{1,r} \sigma_x \right) \left(|j\rangle \langle j+1 | +h.c.\right) , \nonumber \\
    \mathcal{H}_2 &=& \mathcal{H}_{0,2}-\sum_{j=1}^{N-1} \mathrm{i} \left( \Delta_{2,i} \sigma_z  - \Delta_{2,r} \sigma_x \right) \left(|j\rangle \langle j+1 | +h.c.\right) , \nonumber \\
\end{eqnarray}
In principle, one may write 
\begin{equation}
e^{-\mathrm{i} \frac{T}{2} \mathcal{H}_2} e^{-\mathrm{i} \frac{T}{2} \mathcal{H}_1} =e^{-\mathrm{i} T \mathcal{H}_{\rm eff}} ,   
\end{equation}
where $\mathcal{H}_{\rm eff}$ is the effective excitation Hamiltonian that generates the Floquet superoperator. Since $\mathcal{H}_1$ and $\mathcal{H}_2$ are nonHermitian, it follows that $\mathcal{H}_{\rm eff}$ is generally also nonHermitian. 

The above analysis could be easily extended to any general time-periodic Hamiltonian. To this end, we first write the Floquet operator in the form
\begin{eqnarray}
    U_T &=& \lim_{n \rightarrow \infty } e^{-\mathrm{i} H(T) \frac{T}{n}} e^{-\mathrm{i} H\left(\frac{(n-1)T}{n}\right) \frac{T}{n}} \times \cdots \times  e^{-\mathrm{i} H(0) \frac{T}{n}} . \nonumber \\ 
\end{eqnarray}
By considering one exponential at a time and employing Eq.~(\ref{idconj}), we obtain
\begin{equation}
    U_T q_\epsilon U_T^\dagger = u_T q_\epsilon ,
\end{equation}
where 
\begin{equation}
   u_T = \lim_{n \rightarrow \infty } e^{-\mathrm{i} \mathcal{C}_{H(T)} \frac{T}{n}} e^{-\mathrm{i} \mathcal{C}_{H\left(\frac{(n-1)T}{n}\right)} \frac{T}{n}} \times \cdots \times  e^{-\mathrm{i} \mathcal{C}_{H(0)} \frac{T}{n}} .
\end{equation}
Finally, the matrix representation $\mathcal{U}_T$ of the Floquet superoperator $u_T$ could be obtained in the spirit of Eq.~(\ref{mrep}) onwards. More formally, we may write
\begin{equation}
    \mathcal{U}_T = \mathcal{T} e^{-\mathrm{i} \int_0^T \mathcal{H}(t) dt} ,
\end{equation}
where $\mathcal{H}(t)$ is the matrix representation of the commutator superoperator $\mathcal{C}_{H(t)}=[H(t),\cdots]$.

To conclude, given a quadratic time-periodic bosonic Hamiltonian $H(t)$, we can construct the generally non-unitary Floquet superoperator matrix $\mathcal{U}_T$. By writing its eigenvalues as $e^{-\mathrm{i} \epsilon T}$, we can identify $\epsilon$ as the quasienergy excitation. Meanwhile, the elements of the corresponding eigenvector $|q_\epsilon \rangle $ can be used to construct the corresponding quasienergy mode operator $q_\epsilon$.

\section{Floquet modified BKC}
\label{FMBKC}

We consider some periodically driven modified BKC as described by Hamiltonian of the form,
\begin{eqnarray}
    H'(t) &=& \sum_{j=1}^{N-1} \left(J_1(t) a_{A,j}^\dagger a_{B,j} +J_2(t) a_{B,j}^\dagger a_{A,j+1}  \right. \nonumber \\
    &&+ \Delta_1(t) a_{A,j}^\dagger a_{B,j}^\dagger + \left. \Delta_2(t) a_{B,j}^\dagger a_{A,j+1}^\dagger +h.c. \right) ,\nonumber \\ \label{mbkc}
\end{eqnarray}
where $J_1(t)$ and $J_2(t)$ are respectively the (time-periodic) intracell and intercell hopping parameters, $\Delta_1(t)$ and $\Delta_2(t)$ are respectively the (time-periodic) intracell and intercell bosonic analogues of ``pairing" amplitudes, $A$ and $B$ are the two sublattices within a unit cell, and $N$ is the number of cells. We further denote $T$ as the period of $H'(t)$. Unless otherwise stated, we work in the dimensionless units with $T=2$ throughout this paper. 

The static version of Eq.~(\ref{mbkc}) has been intensively studied in Ref.~\cite{Bomantara25}, which uncovers the coexistence of the NHSE and topological zero edge modes. Meanwhile, the time-periodic model of Eq.~(\ref{mbkc}) has yet to be explored in existing literature despite its potential to yield richer properties with no static counterparts. In the following, we aim to close this knowledge gap by analyzing two different time-periodic profiles of the system parameters. For both models, all parameters are subject to binary driving of the form 
\begin{eqnarray}
    J_1(t) &=& \begin{cases}
        J_a & \text{ for } t\;{\rm mod} \;T<T/2 \\
        j_a & \text{ for } t\;{\rm mod} \;T>T/2
    \end{cases} \;,  \nonumber \\
    \Delta_1(t) &=& \begin{cases}
        \Delta_a & \text{ for } t\;{\rm mod} \;T<T/2 \\
        \delta_a & \text{ for } t\;{\rm mod} \;T>T/2
    \end{cases} \;, \nonumber \\
    J_2(t) &=& \begin{cases}
        J_b & \text{ for } t\;{\rm mod} \;T<T/2 \\
        j_b & \text{ for } t\;{\rm mod} \;T>T/2
    \end{cases} \;, \nonumber \\
    \Delta_2(t) &=& \begin{cases}
        \Delta_b & \text{ for } t\;{\rm mod} \;T<T/2 \\
        \delta_b & \text{ for } t\;{\rm mod} \;T>T/2
    \end{cases} \;. \label{bdrive}
\end{eqnarray}
For simplicity, we will further assume that all parameter values are real.

\subsection{Model 1: $j_a=\delta_a=J_b=\Delta_b=0$}
\label{Mod1}

We will further denote the remaining coupling amplitudes in our model 1 as $J_a=J_0$, $\Delta_a=\Delta_0$, $j_b=j_0$, and $\delta_b = \delta_0$. Model 1 represents one of the simplest periodic driving schemes in which the Hamiltonian alternately couples either intracell sites or intercell sites only. As presented below, this model enjoys exact analytical solvability even at general parameter values, thereby providing us clear insight into the interplay among nonHermiticity, periodic driving, and topology in terms of the emergence of NHSE and topological edge modes.

The corresponding matrix representation of the Floquet superoperator can be written in the form of Eq.~(\ref{FloSOm}) with
\begin{eqnarray}
    \mathcal{H}_1 &=& \left(J_0 \sigma_y +\mathrm{i} \Delta_0 \sigma_x \right) \tau_x , \nonumber \\
    \mathcal{H}_2 &=&  \left(j_0 \sigma_y + \mathrm{i}\delta_0 \sigma_x \right) \cos(\hat{k}) \tau_x +(j_0 \sigma_y +\mathrm{i} \delta_0 \sigma_x) \sin(\hat{k}) \tau_y , \nonumber \\
\end{eqnarray}
where $\sigma$ and $\tau$ are Pauli matrices acting, respectively, on the quadrature and sublattice degrees of freedom, and $e^{\pm \mathrm{i} \hat{k}}$ is a translational operator that changes the lattice site $j$ by one unit to $j\pm 1$.

 Under OBC, we define the similarity matrices
\begin{eqnarray}
    A_1' &=& \sum_j \left\lbrace r_1^{-\frac{j+1}{2}} \frac{\sigma_0 +\sigma_z}{2}\frac{\tau_0+\tau_z}{2} + r_1^{-\frac{j}{2}} \frac{\sigma_0 -\sigma_z}{2}\frac{\tau_0-\tau_z}{2} \right. \nonumber \\ \nonumber \\
    &+& \left. r_1^{\frac{j}{2}} \frac{\sigma_0 +\sigma_z}{2}\frac{\tau_0-\tau_z}{2} + r_1^{\frac{j+1}{2}} \frac{\sigma_0 -\sigma_z}{2}\frac{\tau_0+\tau_z}{2} \right\rbrace \otimes |j\rangle \langle j| \nonumber \\
    A_2' &=& \sum_j \left\lbrace r_2^{-\frac{j}{2}} \frac{\sigma_0 -\sigma_z}{2}\frac{\tau_0+\tau_z}{2} + r_2^{-\frac{j}{2}} \frac{\sigma_0 +\sigma_z}{2}\frac{\tau_0-\tau_z}{2} \right. \nonumber \\ \nonumber \\
    &+& \left. r_2^{\frac{j}{2}} \frac{\sigma_0 +\sigma_z}{2}\frac{\tau_0+\tau_z}{2} + r_2^{\frac{j}{2}} \frac{\sigma_0 -\sigma_z}{2}\frac{\tau_0-\tau_z}{2} \right\rbrace \otimes |j\rangle \langle j| , \nonumber \\
\end{eqnarray}
where $r_1=\frac{\Delta_0+J_0}{\Delta_0 -J_0}$ and $r_2=\frac{\delta_0+j_0}{\delta_0 -j_0}$. It follows that \cite{Bomantara25}
\begin{eqnarray}
    A^{-1} \mathcal{H}_1 A &=& \tilde{D}_0 \sigma_x \tau_x , \nonumber \\
    A^{-1} \mathcal{H}_2 A &=& \tilde{d}_0 \sigma_x (\cos(\hat{k}) \tau_x +\sin(\hat{k}) \tau_y) , 
\end{eqnarray}
where $A=A_2'A_1'$ and
\begin{eqnarray}
    \tilde{D}_0  &=& \begin{cases}
        \mathrm{i} \sqrt{\Delta_0^2 -J_0^2}  & \text{ for } \Delta_0 > J_0 \\
        \sqrt{J_0^2 -\Delta_0^2} & \text{ for } J_0 > \Delta_0 \\ 
    \end{cases} , \nonumber \\
    \tilde{d}_0 &=& \begin{cases}
        \mathrm{i} \sqrt{\delta_0^2 -j_0^2}  & \text{ for } \delta_0 > j_0 \\
        \sqrt{j_0^2 -\delta_0^2} & \text{ for } j_0 > \delta_0 \\
    \end{cases} . \nonumber \\
\end{eqnarray}
The similarity transformation then yields the following form for the Floquet superoperator under OBC,
\begin{equation}
   \tilde{\mathcal{U}}_T \equiv A^{-1} \mathcal{U}_T A = e^{-\mathrm{i} \frac{\tilde{d}_0 T}{2} \sigma_x (\cos(\hat{k}) \tau_x +\sin(\hat{k}) \tau_y)} e^{-\mathrm{i} \frac{\tilde{D}_0 T}{2} \sigma_x \tau_x} ,
\end{equation}
In particular, $\tilde{\mathcal{U}}_T$ is unitary for $J_0>\Delta_0$ and $j_0>\delta_0$, the quasienergy excitations of which are completely real. Meanwhile, under PBC, the same similarity transformation above does not yield a unitary Floquet superoperator under $J_0>\Delta_0$ and $j_0>\delta_0$ due to the presence of hopping and ``pairing" connecting the two end points. Consequently, the system's quasienergy excitation spectrum depends sensitively on the choice of boundary conditions, thereby signifying the presence of NHSE \cite{Yao18}.  

\begin{center}
    \begin{figure}
        \centering
        \includegraphics[scale=0.35]{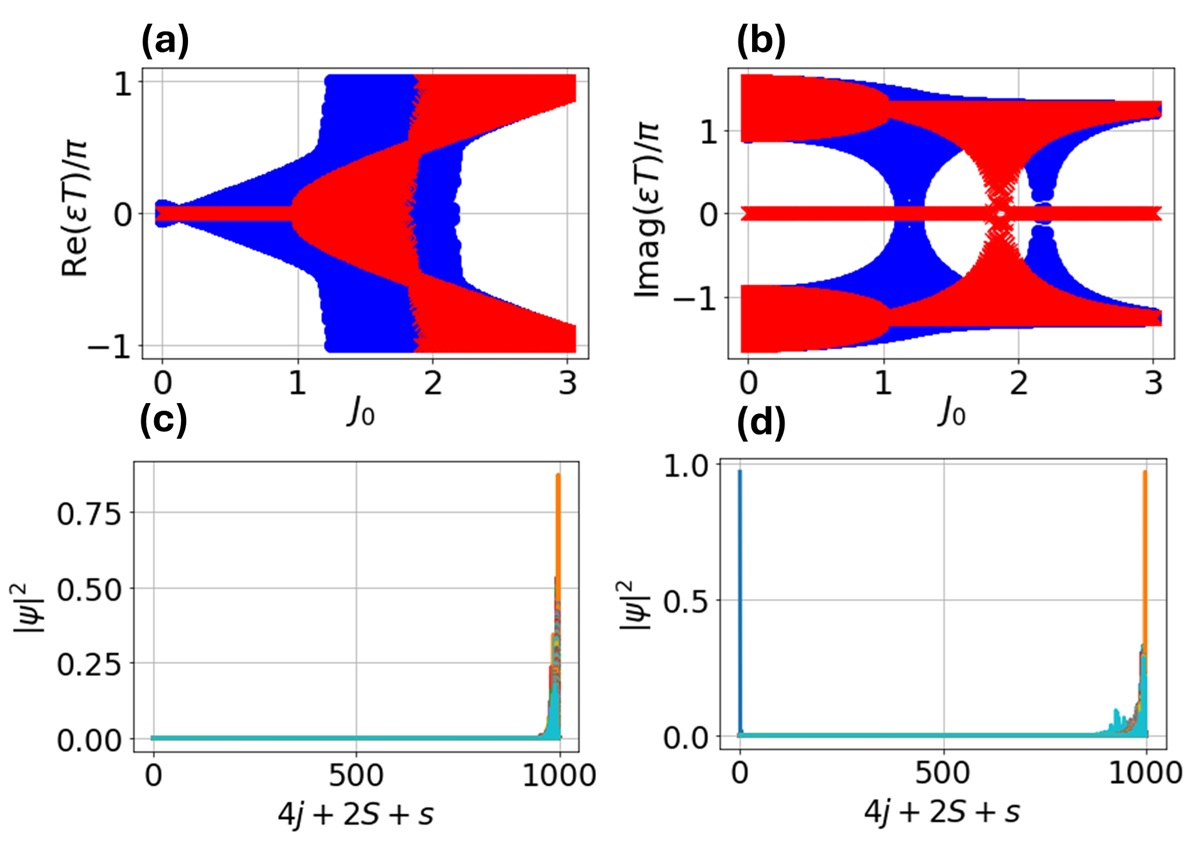}
        \caption{(a,b) The real and imaginary parts of the system's quasienergy excitation spectrum under OBC (red) and PBC (blue). (c,d) The spatial profiles of all quasienergy excitation eigenstates under (c) $J_0=0.5$ and (d) $J_0=3$. The other system parameters are taken as $j_0=0.5$, $\Delta_0=1$, $\delta_0=4$, (a,b) $N=100$, (c,d) $N=250$.}
        \label{fig:fse}
    \end{figure}
\end{center}

In Fig.~\ref{fig:fse}(a,b), we numerically demonstrate NHSE at general parameter values by computing the quasienergy excitation spectra under OBC and PBC side-by-side, which highlights their difference. Moreover, we further confirm the localization of all quasienergy excitation eigenstates near a system's edge by computing their spatial profiles in Fig.~\ref{fig:fse}(c,d). Here, the spatial profile of a quasienergy excitation eigenstate $|\psi\rangle$ is defined as 
\begin{equation}
    |\psi|^2(4j+2S+s)=\langle j,S,s | \psi \rangle , \label{sprof}
\end{equation}
where $j$ is the site number, $S=0$ ($1$) for sublattice A (B), and $s=0$ ($1$) in the $x$ ($p$) subspace. The combination $4j+S+s$ is chosen so that a unique positive integer is uniquely assigned for each $(j,S,s)$. Under the parameter values considered in Fig.~\ref{fig:fse}(c), we observe that all eigenstates are localized near the right end. Meanwhile, under a different set of parameter values (Fig.~\ref{fig:fse}(d)), a single quasienergy excitation eigenstate is localized near the left end, while all other eigenstates are localized near the right end. We find that the only left-edge-localized eigenstate of Fig.~\ref{fig:fse}(d) corresponds to $\varepsilon =\pi/T$, i.e., the so-called topological $\pi$ modes that have no static counterparts.  

\begin{center}
    \begin{figure*}
        \includegraphics[scale=0.41]{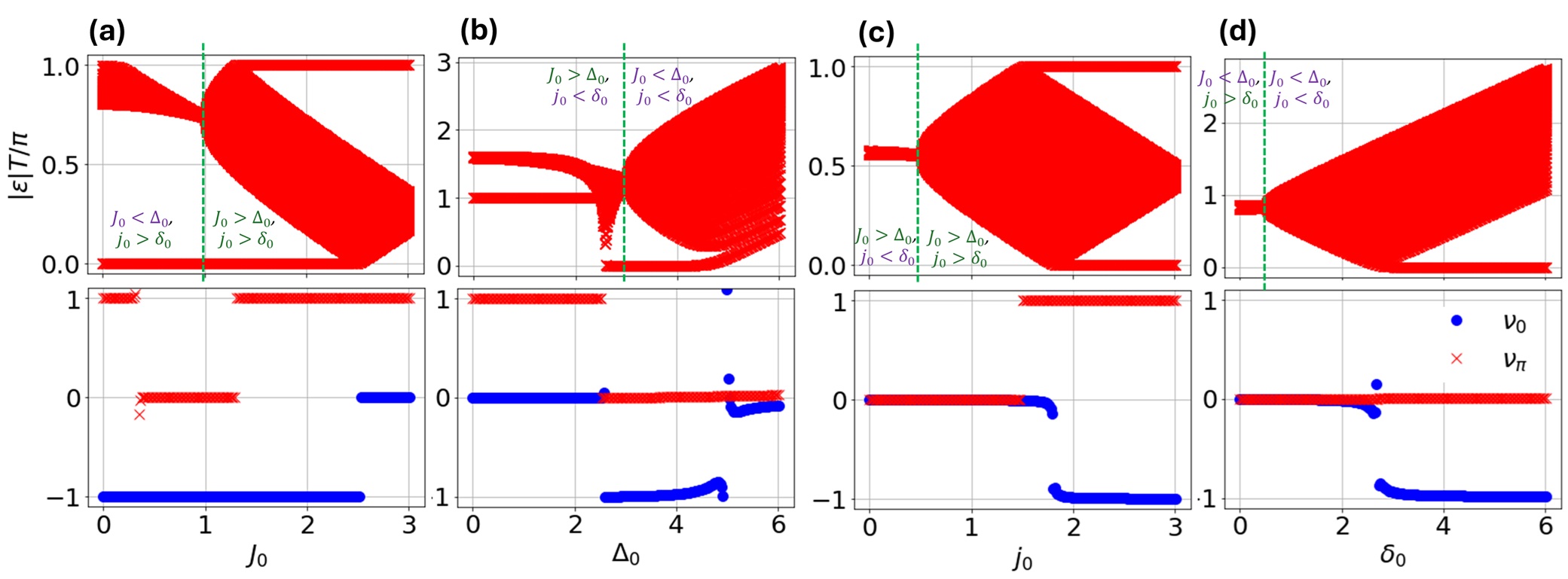}
        \caption{The upper panels depict the system's quasienergy excitation spectrum for model 1 under OBC as a system parameter is varied. The bottom panels present the numerically calculated winding numbers $\nu_0$ and $\nu_\pi$. The remaining system parameters are chosen as $N=100$, (a) $\Delta_0 = 1$, $\delta_0  = 0.5$, $j_0 = 4$, (b) $J_0 = 3$, $j_0  = 0.5$, $\delta_0 = 4$, (c) $J_0 =2$, $\Delta_0 =1$, $\delta_0 =0.5$, (d) $J_0 =3$, $j_0 =0.5$, $\Delta_0 =4$. }
        \label{fig:ftem}
    \end{figure*}
\end{center}

To further elucidate the presence and key features of the topological edge modes in the system, we numerically plot $|\varepsilon |$ versus each system parameter in the upper panels Fig.~\ref{fig:ftem}. The remaining system parameters are chosen so that the four upper panels of Fig.~\ref{fig:ftem} capture all possible cases of $J_0<\Delta_0$ and/or $j_0<\delta_0$, as well as $J_0>\Delta_0$ and/or $j_0>\delta_0$. There, topological zero and $\pi$ modes are clearly observed in some regions following the closing and reopening of bulk bands at $\varepsilon=0$ and $\varepsilon= \pm \pi/T$ respectively. 

We now adapt the construction of Ref.~\cite{Asboth14} to develop a set of topological invariants associated with the above obtained zero and $\pi$ modes. To this end, we further transform $\tilde{\mathcal{U}}_T$ to the ``symmetric-time-frame" form
\begin{equation}
    \tilde{\mathcal{U}}_T \rightarrow F G , 
\end{equation}
where 
\begin{eqnarray}
    G= e^{-\mathrm{i} \frac{\tilde{D}_0 T}{4} \sigma_x \tau_x} e^{-\mathrm{i} \frac{\tilde{d}_0 T}{4} \sigma_x (\cos(\hat{k}) \tau_x +\sin(\hat{k}) \tau_y)} &,& G = \tau_z F^{-1} \tau_z . \nonumber \\ 
\end{eqnarray}
In this case $\tau_z$ is identified as the chiral symmetry of the system which satisfies $\tau_z \tilde{\mathcal{U}}_T\tau_z = \tilde{\mathcal{U}}_T^{-1}$. In the canonical basis of $\tau_z ={\rm diag}(1,-1)$, $G$ can be written as the block matrix
\begin{equation}
    G = \left( \begin{array}{cc}
        A & B \\
        C & D 
    \end{array} \right),
\end{equation}
where $A$, $B$, $C$, and $D$ are some $2\times 2$ matrices. The winding numbers characterizing the number of topological zero and $\pi$ modes could then be defined as \cite{Bomantara18b},
\begin{eqnarray}
    \nu_0 = \frac{1}{4\pi \mathrm{i}}\oint \mathrm{Tr}\left(B^{-1} dB\right) &,& \nu_\pi = \frac{1}{4\pi \mathrm{i}}\oint \mathrm{Tr}\left(D^{-1} dD\right) .\nonumber \\
\end{eqnarray}
We have numerically computed $\nu_0$ and $\nu_\pi$ in the bottom panels of Fig.~\ref{fig:ftem} to confirm that they correctly predict the respective presence of topological zero and $\pi$ modes. 

To gain analytical insight into the formation of topological zero and $\pi$ modes, we first note that the above expressions can further be reduced to the simple forms 
\begin{eqnarray}
    \nu_0 &=& \frac{1}{2\pi \mathrm{i}} \oint_{\mathcal{C}_0}\frac{1}{z_0 +z} dz, \nonumber \\
    \nu_\pi &=& \frac{1}{2\pi \mathrm{i}} \oint_{\mathcal{C}_\pi}\frac{1}{z_\pi +z} dz , \label{fwn}
\end{eqnarray}
where $\mathcal{C}_0$ and $\mathcal{C}_\pi$ are circular contours of radii (respectively) $r_0$ and $r_\pi$ around the origin in the complex plane and, using the convention $\sqrt{-1}=\mathrm{i}$ if applicable,
\begin{eqnarray}
    z_0 &=& \sin\left(\frac{\sqrt{J_0^2-\Delta_0^2} T}{4}\right) \cos\left(\frac{\sqrt{j_0^2-\delta_0^2} T}{4}\right) , \nonumber \\
    r_0 &=& \cos\left(\frac{\sqrt{J_0^2-\Delta_0^2} T}{4}\right) \sin\left(\frac{\sqrt{j_0^2-\delta_0^2} T}{4}\right) , \nonumber \\
    z_\pi &=& -\cos\left(\frac{\sqrt{J_0^2-\Delta_0^2} T}{4}\right) \cos\left(\frac{\sqrt{j_0^2-\delta_0^2} T}{4}\right) , \nonumber \\
    r_\pi &=& \sin\left(\frac{\sqrt{J_0^2-\Delta_0^2} T}{4}\right) \sin\left(\frac{\sqrt{j_0^2-\delta_0^2} T}{4}\right) .
\end{eqnarray}
The residue theorem then quickly yields
\begin{eqnarray}
    \nu_0 =\begin{cases}
        \pm 1 & \text{ for } |z_0| <r_0 \\
        0 & \text{ for } |z_0|>r_0
    \end{cases} &,& \nu_\pi =\begin{cases}
        \pm 1 & \text{ for } |z_\pi| <r_\pi \\
        0 & \text{ for } |z_\pi| >r_\pi
    \end{cases} .\nonumber \\ \label{wres}
\end{eqnarray}
For $J_0>\Delta_0$ and $j_0>\delta_0$ ($\tilde{\mathcal{U}}_T$ is unitary), this translates to $\left(\frac{\sqrt{J_0^2-\Delta_0^2} T}{2}<\frac{\sqrt{j_0^2-\delta_0^2} T}{2}\right)\; {\rm mod}\; 2\pi$ and $\left(\frac{\sqrt{J_0^2-\Delta_0^2} T}{2}>\pi-\frac{\sqrt{j_0^2-\delta_0^2} T}{2}\right) \; {\rm mod}\; 2\pi$, with $n\in \mathbb{Z}$, for the the emergence of topological zero modes and $\pi$ modes respectively. These conditions agree with the well-established results for a typical Hermitian Floquet SSH model \cite{Tan20} upon the identification that $\frac{\sqrt{J_0^2-\Delta_0^2} T}{2}$ represents the intracell hopping, whereas $\frac{\sqrt{j_0^2-\delta_0^2} T}{2}$ represents the intercell hopping.

For $J_0<\Delta_0$ and $j_0>\delta_0$, Eq.~(\ref{wres}) yields the conditions $\tanh^2\left(\frac{\sqrt{\Delta_0^2-J_0^2} T}{4}\right) < \tan^2 \left(\frac{\sqrt{j_0^2-\delta_0^2} T}{4}\right)$ and $\tanh^2\left(\frac{\sqrt{\Delta_0^2-J_0^2} T}{4}\right)>\cot^2 \left(\frac{\sqrt{j_0^2-\delta_0^2} T}{4}\right)$ for the existence of topological zero modes and $\pi$ modes respectively. Given that the hyperbolic tangent function is always less than 1, $\pi$ modes \emph{cannot} exist in the regime $(4n-1)\pi<\sqrt{j_0^2-\delta_0^2} T<(4n+1)\pi$ with $n\in\mathbb{Z}$. Interestingly, at $\sqrt{j_0^2-\delta_0^2} T = (4n+2)\pi$, topological zero and $\pi$ modes always coexist regardless of $J_0$ and $\Delta_0$. 

For $J_0>\Delta_0$ and $j_0<\delta_0$, Eq.~(\ref{wres}) implies that zero and $\pi$ modes exist whenever, respectively, $\tan^2\left(\frac{\sqrt{J_0^2-\Delta_0^2} T}{4}\right) < \tanh^2 \left(\frac{\sqrt{\delta_0^2-j_0^2} T}{4}\right)$ and $\cot^2\left(\frac{\sqrt{J_0^2-\Delta_0^2} T}{4}\right) < \tanh^2 \left(\frac{\sqrt{\delta_0^2-j_0^2} T}{4}\right)$. As the two conditions cannot be simultaneously satisfied unless at a very specific value of $\sqrt{J_0^2-\Delta_0^2} T=(4n+1)\pi$ with $n\in \mathbb{Z}$, topological zero and $\pi$ modes generally cannot coexist.

Finally, for $J_0<\Delta_0$ and $j_0<\delta_0$, the condition for the existence of topological zero modes is identical to that in a static modified BKC model \cite{Bomantara25}, i.e., $\sqrt{\delta_0^2-j_0^2} > \sqrt{\Delta_0^2-J_0^2}$. On the other hand, the existence of topological $\pi$ modes demands the condition $\tanh^2 \left(\frac{\sqrt{\delta_0^2-j_0^2} T}{4}\right)>\coth^2 \left(\frac{\sqrt{\Delta_0^2-J_0^2} T}{4}\right)$, which can never be satisfied as hyperbolic tangent is always smaller than the hyperbolic cotangent. Consequently, topological $\pi$ modes cannot exist in the regime $J_0<\Delta_0$ and $j_0<\delta_0$. Intuitively, this result could be understood from the fact that the quasienergy excitation spectrum is purely imaginary for $J_0<\Delta_0$ and $j_0<\delta_0$, which is not periodic modulo the driving frequency like its real component. Consequently, these purely imaginary quasienergy excitations behaves in exactly the same manner as imaginary energy excitations in a typical static system and cannot support topological edge modes beyond those found in its static counterpart. 

\subsection{Model 2: $\mathrm{sgn}(J_a j_a)=\mathrm{sgn}(\Delta_a \delta_a)=-\mathrm{sgn}(J_b j_b)=-\mathrm{sgn}(\Delta_b \delta_b)$ }
\label{Mod2}

In model 2, all parameter values are generally nonzero at any instant, but we require that there is a relative sign between the intracell and intercell couplings over the duration of half a period, while there is no relative sign between the two types of couplings over the other half of the period. The motivation for considering this model is that it has the potential to generate any large number of topological zero and $\pi$ modes in a controllable manner. 

For further simplification of the model, we shall further set
\begin{equation}
    -\frac{j_a}{J_a}=\frac{j_b}{J_b}=-\frac{\delta_a}{\Delta_a}=\frac{\delta_b}{\Delta_b}=m ,
\end{equation}
where $m \in \mathbb{R}$. In this case, the same similarity transformation with respect to $A=A_2' A_1'$ as in Sec.~\ref{Mod1} could be applied to yield 
\begin{equation}
    \tilde{\mathcal{U}}_T \equiv A^{-1} \mathcal{U}_T A = e^{-\frac{\tilde{\mathcal{H}}_2T}{2}} e^{-\frac{\tilde{\mathcal{H}}_1T}{2}} , \label{mod2trans}
\end{equation}
where 
\begin{eqnarray}
    \tilde{\mathcal{H}}_1 &=& \tilde{D}_a \sigma_x \tau_x + \tilde{D}_b \sigma_x \left( \cos(\hat{k}) \tau_x + \sin(\hat{k}) \tau_y \right) ,\nonumber \\
    \tilde{\mathcal{H}}_2 &=& -m\tilde{D}_a \sigma_x \tau_x + m\tilde{D}_b \sigma_x \left( \cos(\hat{k}) \tau_x + \sin(\hat{k}) \tau_y \right)  , \nonumber \\
    \tilde{D}_\ell &=& \begin{cases}
        \mathrm{i} \sqrt{\Delta_\ell^2 -J_\ell^2}  & \text{ for } \Delta_\ell > J_\ell \\
        \sqrt{J_\ell^2 -\Delta_\ell^2} & \text{ for } J_\ell > \Delta_\ell \\ 
    \end{cases} .  \label{defD}
\end{eqnarray}

\begin{center}
    \begin{figure*}
        \includegraphics[scale=0.4]{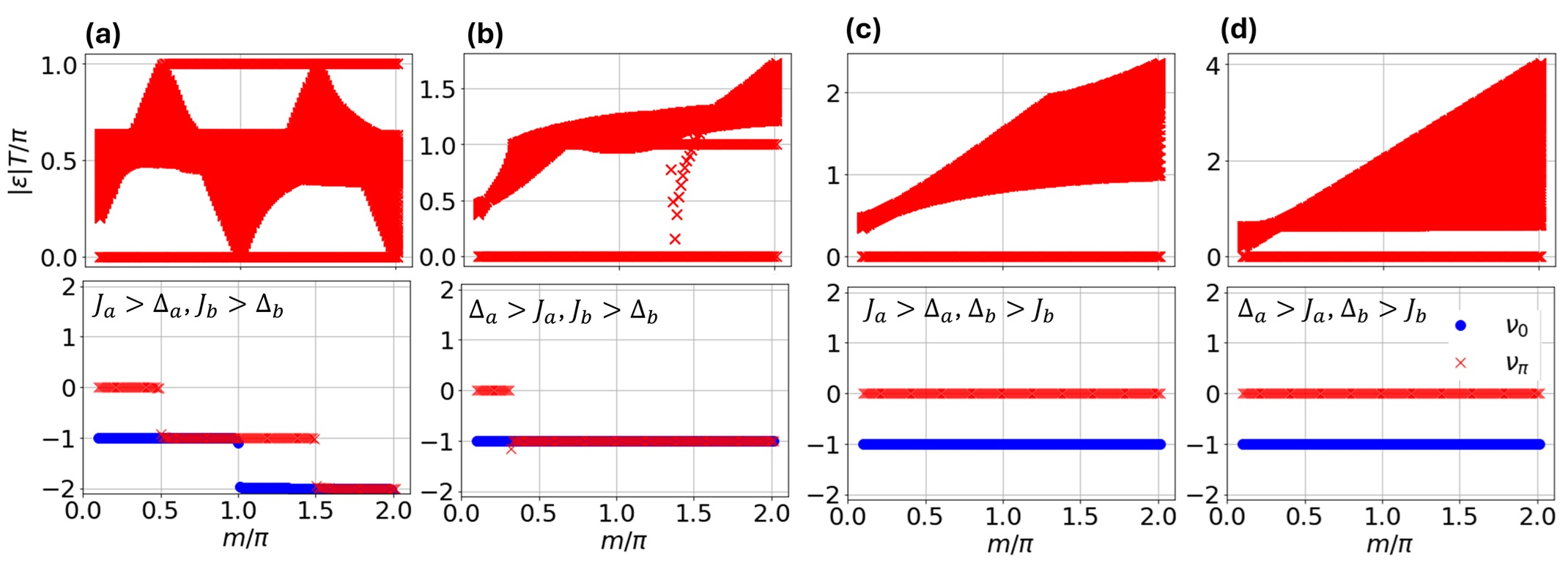}
        \caption{The upper panels depict the system's quasienergy excitation spectrum for model 2 under OBC as a system parameter is varied according to case(i) in Sec.~\ref{Mod2}. The bottom panels present the numerically calculated winding numbers $\nu_0$ and $\nu_\pi$. The remaining system parameters are chosen as $N=100$, (a) $\Delta_a = 0.6$, $\Delta_b=1.6$, $J_a  = \sqrt{\Delta_a^2 +1}$, $J_b  = \sqrt{\Delta_b^2 +1}$, (b) $J_a = 0.6$, $\Delta_b=1.6$, $\Delta_a  = \sqrt{J_a^2 +1}$, $J_b  = \sqrt{\Delta_b^2 +1}$, (c) $\Delta_a = 0.6$, $J_b=1.6$, $J_a  = \sqrt{\Delta_a^2 +1}$, $\Delta_b  = \sqrt{J_b^2 +1}$, (d) $J_a = 0.6$, $J_b=1.6$, $\Delta_a  = \sqrt{J_a^2 +1}$, $\Delta_b  = \sqrt{J_b^2 +1}$. }
        \label{fig:ftem2}
    \end{figure*}
\end{center}

When $J_a>\Delta_a$ and $J_b>\Delta_b$, both $\tilde{\mathcal{H}}_1$ and $\tilde{\mathcal{H}}_2$ reduce to Hermitian SSH matrices. The resulting unitary $\tilde{\mathcal{U}}_T$, which is mathematically similar to the Floquet operator of the model studied in Ref.~\cite{Bomantara20c}, has the remarkable property that it can support any desired number of topological zero and $\pi$ modes that is controlled by the tunable parameter $m$. Specifically, in the special case of $\sqrt{J_a^2-\Delta_a^2}=\sqrt{J_b^2-\Delta_b^2}=1$ and $\frac{\ell\pi}{2}<m<\frac{(\ell+1)\pi}{2}$, there are in total $\ell+1$ ($\ell$) pairs of zero modes and $\ell$ pairs of $\pi$ modes if $\ell$ is an even (odd) integer \cite{Bomantara20c}. In Fig.~\ref{fig:ftem2}(a), we numerically confirm this by plotting the system's quasienergy excitation spectrum, along with the winding numbers $\nu_0$ and $\nu_\pi$ defined previously in Sec.~\ref{Mod1}.  

When $J_a<\Delta_a$ and/or $J_b<\Delta_b$, our numerical results in Fig.~\ref{fig:ftem2}(b-d) reveal the absence of periodic band touching and reopening events as $m$ is varied. Consequently, the number of topological zero and $\pi$ modes does not increase indefinitely with $m$. This finding is hardly surprising considering that the proliferation of the topological edge modes results from the periodicity in the real part of the quasienergy, which in turn allows the systematic formation of band touching and reopening events responsible for the regular topological phase transitions. As the quasienergy becomes complex, as is the case when $J_a<\Delta_a$ and/or $J_b<\Delta_b$, the unbounded nature of its imaginary part makes it harder for the band touching and reopening events to occur. Nevertheless, as long as $J_b>\Delta_b$, topological zero and $\pi$ modes may still be present and coexist in some parameter regime (Fig.~\ref{fig:ftem2}(b)).   

%\vspace{0.5cm}

%\noindent {\it Case (ii):} $\frac{j_a}{\Delta_a}=-\frac{j_b}{J_b}=\frac{\delta_a}{J_a}=-\frac{\delta_b}{\Delta_b}=m$, $m\in \mathbb{R}$. \newline

%In this case, the same similarity transformation as before again yields Eq.~(\ref{mod2trans}), but with
%\begin{eqnarray}
%    \tilde{\mathcal{H}}_1 &=& \tilde{D}_1 \sigma_x \tau_x + \tilde{D}_2 \sigma_x \left( \cos(\hat{k}) \tau_x + \sin(\hat{k}) \tau_y \right) ,\nonumber \\
%    \tilde{\mathcal{H}}_2 &=& \mathrm{i} m\tilde{D}_1 \sigma_y \tau_x + m\tilde{D}_2 \sigma_x \left( \cos(\hat{k}) \tau_x + \sin(\hat{k}) \tau_y \right)  , \nonumber \\
%\end{eqnarray}
%where $\tilde{D}_\ell$ is as defined in case (i)...

\section{Discussion}

\subsection{Effect of onsite bosonic frequency}
\label{onsite}

The BKC \cite{McDonald18} and its dimerized generalization \cite{Bomantara25} are known to be very sensitive to the presence of finite onsite bosonic frequency, corresponding to an additional term of the form
\begin{equation}
    H_\omega = \sum_{j=1}^{N} \omega a_j^\dagger a_j . \label{om}
\end{equation}
In particular, such a term is found to destroy NHSE regardless of how small $\omega$ is. Moreover, while some topological edge modes persist in the presence of such a term, others quickly disappear as $\omega$ becomes nonzero. It is of course worth pointing out that the detrimental effect of finite onsite bosonic frequency does \emph{not} pose a real challenge in experiments, as such an effect could be easily eliminated, e.g., by working in the rotating frame \cite{McDonald18,Bomantara25}. Nevertheless, exploring the effect of onsite bosonic frequency is still of fundamental interest, as it yields better understanding on the behavior of nonHermitian systems. For this reason, this section is dedicated to uncovering the fate of the nonHermitian physics observed in the two periodically driven modified BKC models introduced in the previous section in the presence of finite onsite bosonic frequency. 

\begin{center}
    \begin{figure}
        \centering
        \includegraphics[scale=0.35]{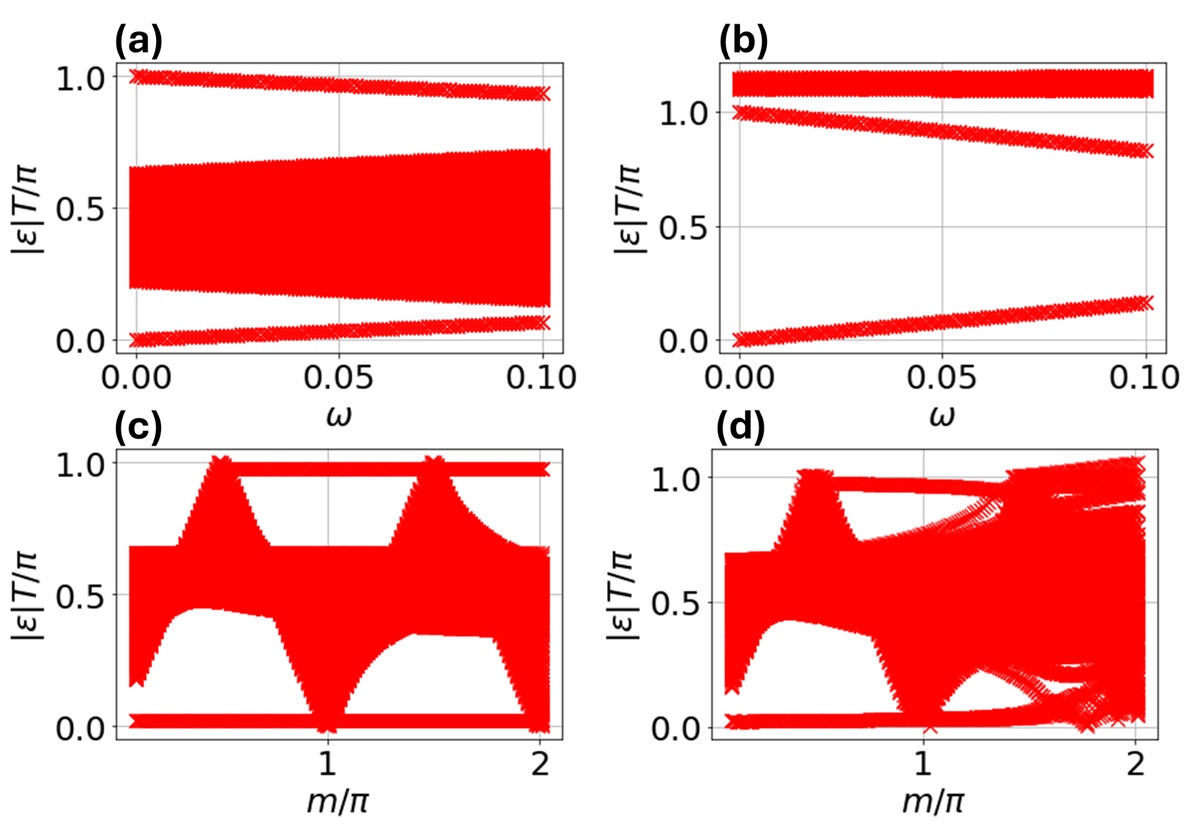}
        \caption{The system's quasienergy excitation spectrum under OBC at finite onsite bosonic frequency for (a,b) model 1 and (c,d) model 2. (a,b) Model 1 system parameters are chosen as (a) $\Delta_0=0.5$ $\delta_0=1$, $J_0=\sqrt{1.8^2+\Delta_0^2}$, and $j_0=\sqrt{2.5^2+\delta_0^2}$, (b) $J_0=0.5$, $\delta_0=1$, $\Delta_0=\sqrt{1.8^2+J_0^2}$, and $j_0=\sqrt{(\pi-0.2)^2+\delta_0^2}$. (c,d) Model 2 system parameters are chosen as (c) $\Delta_a=-\frac{\delta_a}{m}=0.6$, $\Delta_b=\frac{\delta_b}{m}=0.6$, $J_a=-\frac{j_a}{m}=\sqrt{1^2+0.6^2}$, and $J_b=\frac{j_b}{m}=\sqrt{1^2+0.6^2}$, (d) $\Delta_a=-\frac{\delta_a}{m}=1$, $\Delta_b=\frac{\delta_b}{m}=0.6$, $J_a=-\frac{j_a}{m}=\sqrt{2}$, and $J_b=\frac{j_b}{m}=\sqrt{1^2+0.6^2}$. The system size is taken as $N=100$ in all panels, and $\omega=0.01\pi$ in panels (c,d).}
        \label{fig:finw}
    \end{figure}
\end{center}

We start by adding $H_\omega$ (Eq.~(\ref{om})) to model 1 of Sec.~\ref{Mod1}. By focusing on two different sets of parameter values that support both topological zero and $\pi$ modes in the $\omega\rightarrow 0$, we investigate the transformation of the quasienergy spectra as $\omega$ is varied. Our results are summarized in Fig.~\ref{fig:finw}(a,b), where the system parameters satisfy $J_0>\Delta_0$ and $j_0>\delta_0$ for panel (a), whereas $J_0<\Delta_0$ and $j_0>\delta_0$ are chosen for panel (b). In both cases, topological zero and $\pi$ modes are observed to demonstrate some robustness against finite onsite bosonic energy, i.e., the zero and $\pi/T$ quasienergy excitations at $\omega=0$ form continuous curves at small to moderate $\omega$, although they are no longer pinned at zero and $\pi/T$ quasienergy excitations respectively due to the breakdown of chiral symmetry.    

In Fig.~\ref{fig:finw}(c,d), we investigate the effect of $H_\omega$ on the formation of multiple topological phase transitions in model 2 of Sec.~\ref{Mod2} as the ratio parameter $m$ is varied. While different sets of system parameters are chosen for both panels, they satisfy $\tilde{D}_a=\tilde{D}_b=1$ (Recall Eq.~(\ref{defD}) for their definitions). At $\omega=0$, both panels yield the same Floquet superoperator matrix under appropriate similarity transformation, i.e., Eq.~(\ref{defD}), and consequently have identical quasienergy excitation spectrum. As Fig.~\ref{fig:finw}(c,d) demonstrates, however, the presence of $H_\omega$ causes the quasienergy excitation spectrum to also depend on the individual parameters $J_a$, $J_b$, $\Delta_a$, and $\Delta_b$, resulting in both panels exhibiting significantly different profiles. In particular, a qualitatively similar quasienergy excitation spectrum as its $\omega=0$ counterpart is observed when $J_a=J_b$ and $\Delta_a=\Delta_b$ (Fig.~\ref{fig:finw}(c)). When $J_a\neq J_b$ and $\Delta_a\neq \Delta_b$ (Fig.~\ref{fig:finw}(d)), while some topological phase transitions are still observed at small $m$, the quasienergy excitation spectrum becomes gapless at a sufficiently large value of $m$, resulting in the disappearance of all topological edge modes.         

\begin{center}
    \begin{figure}
        \centering
        \includegraphics[scale=0.35]{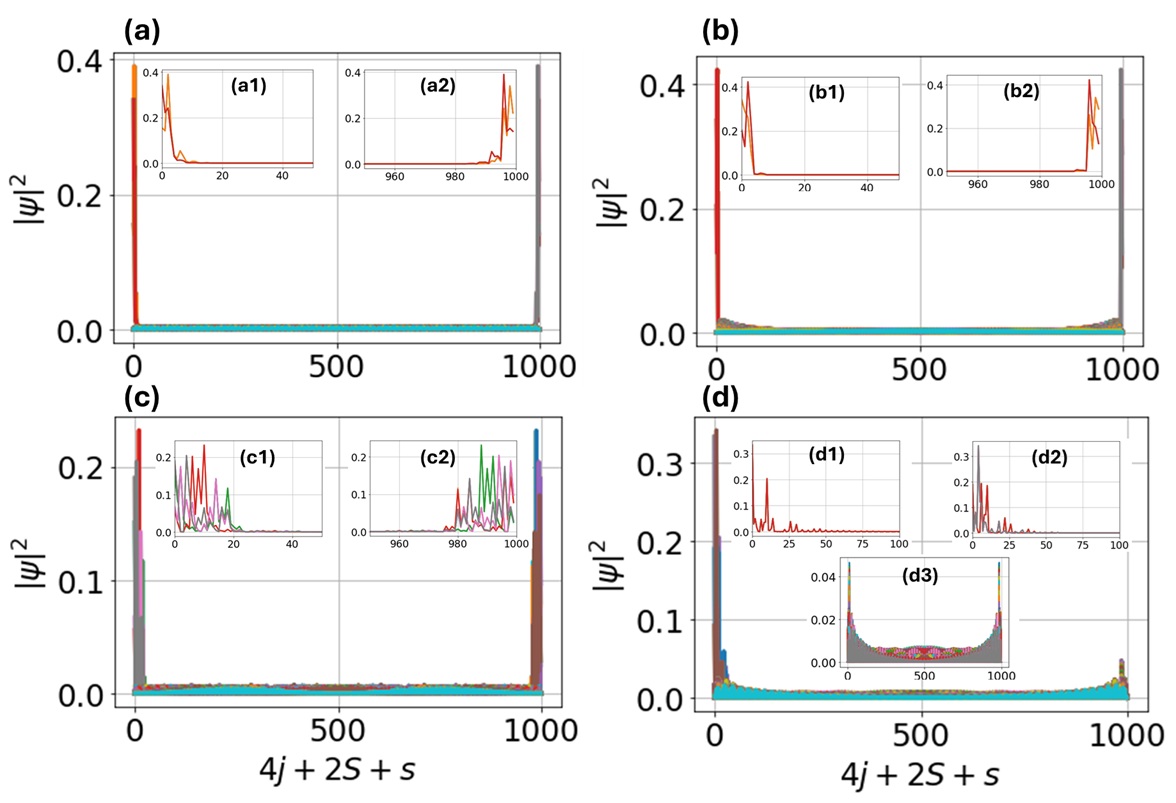}
        \caption{The spatial profiles of all quasienergy excitation eigenstates (see Eq.~(\ref{sprof}) for its definition) under (a,b) model 1, (c,d) model 2. The insets of each panel depict the zoomed-in view of all localized quasienergy excitation eigenstates, which include nontopological bulk modes (panel (d3)) and genuine topological edge modes (all the other panels). The system parameters are taken as $N=250$ in all panels, $\omega=0.05$ in panels (a,b), $m=1.8\pi$ in panel (c), and $m=1.1\pi$ in panel (d). The remaining system parameters respectively match those of Fig.~\ref{fig:finw}(a)-(d).}
        \label{fig:fmodwfw}
    \end{figure}
\end{center}

For completeness, we have also plotted the spatial profiles of all quasienergy excitation eigenstates of both models under different sets of parameter values in Fig.~\ref{fig:fmodwfw}. In all panels, we observe that all bulk states generally become delocalized; any observed localized profiles correspond to a quasienergy excitation close to either $0$ or $\pm \pi/T$, thereby representing zero or $\pi$ modes, respectively. In particular, we have confirmed that there are precisely only two zero modes and two $\pi$ modes localized at each end of the chain for model 1 under both sets of parameter values, the spatial profiles of which are plotted separately in the insets of Fig.~\ref{fig:fmodwfw}(a,b). These numbers agree with the numbers of zero and $\pi$ modes achievable in model 1 at $\omega=0$, as elucidated in Sec.~\ref{Mod1}. 

For model 2, we find that there are exactly four zero modes and four $\pi$ modes localized at each end of the chain in Fig.~\ref{fig:fmodwfw}(c) (see its insets for a closer inspection of their spatial profiles). These numbers again match the expected numbers of zero and $\pi$ modes at $\omega=0$, demonstrating that each topological phase transition observed in Fig.~\ref{fig:finw}(c) indeed alters the number of zero/$\pi$ modes by two pairs. Interestingly, while four pairs of zero modes and two pairs of $\pi$ modes are expected and also found in Fig.~\ref{fig:fmodwfw}(d), \emph{all} of them are localized near the left end of the lattice. Moreover, while the corresponding bulk states are generally considered delocalized, i.e., they have significant support on all sites throughout the bulk, significant peaks are clearly observed near the left and right ends of the lattice, suggesting a faint remnant of the NHSE.   

\begin{center}
    \begin{figure}
        \centering
        \includegraphics[scale=0.35]{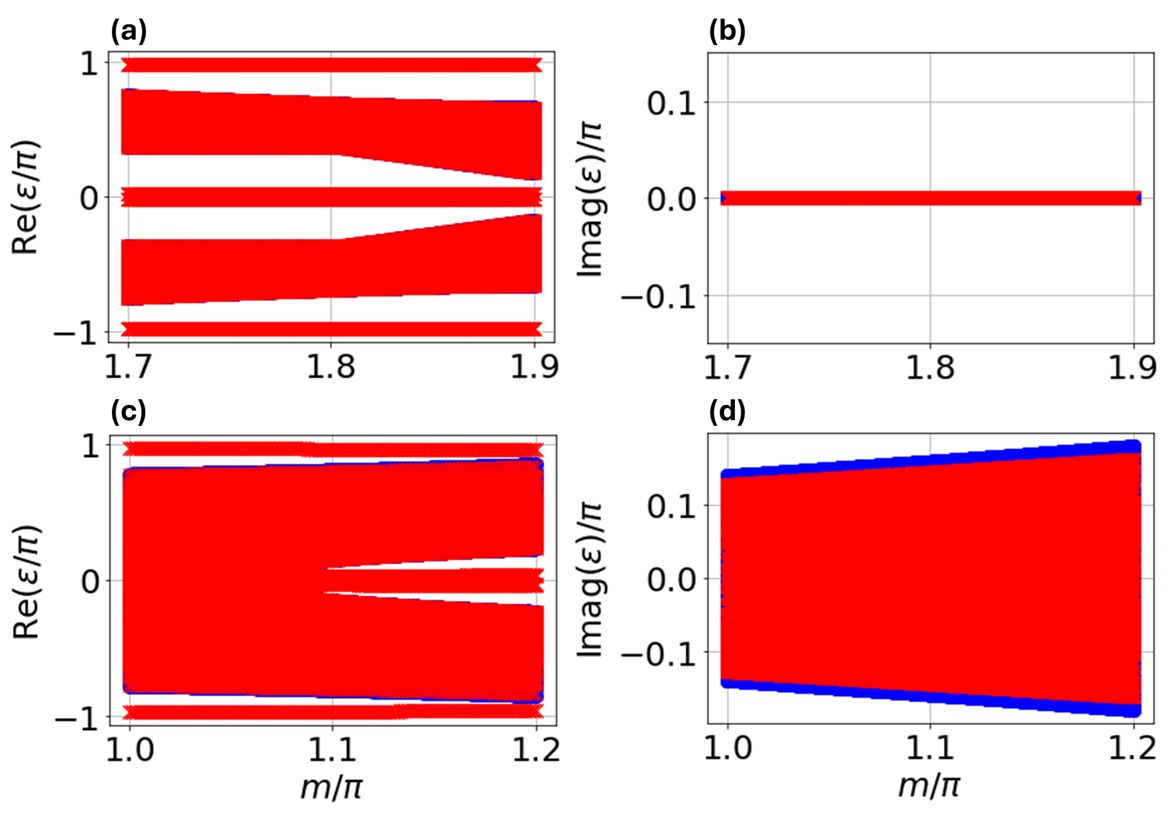}
        \caption{The real and imaginary parts of the system's quasienergy excitation spectrum under OBC (red) and PBC (blue). All system parameters are chosen to be the same as those in (a,b) Fig.~\ref{fig:finw}(c) and (c,d) Fig.~\ref{fig:finw}(d). }
        \label{fig:fmodbkcnhsew}
    \end{figure}
\end{center}

By plotting the real and imaginary parts of the quasienergy excitations in Fig.~\ref{fig:fmodbkcnhsew}, we observe identical PBC and OBC spectra in the vicinity of the parameters of Fig.~\ref{fig:finw}(c), hence confirming the absence of NHSE (Fig.~\ref{fig:fmodbkcnhsew}(a,b)). Meanwhile, in the vicinity of the parameters of Fig.~\ref{fig:finw}(d), the PBC and OBC spectra differ slightly, which signifies that NHSE remains (Fig.~\ref{fig:fmodbkcnhsew}(c,d)). It is worth noting that the preservation of NHSE at $\omega\neq 0$ is a distinctive feature of a periodically driven system that has no static counterpart.  

\subsection{Effect of spatial disorder}
\label{disorder}

Topological edge modes are expected to be robust in the presence of spatial disorder, provided the protecting symmetry is preserved and that the quasienergy bulk gap remains. Both models of our Floquet modified BKC are protected by the chiral symmetry, which is respected whenever $\omega=0$. It is easily verified that applying disorder on any of the system parameters does not break the chiral symmetry. However, it remains to be checked if the presence of disorder leads to accidental quasienergy gap closing. Therefore, we shall now explicitly compute the quasienergy spectra in the presence of various spatial disorders to directly verify the robustness of our systems' topology.   

\begin{center}
    \begin{figure*}
        \includegraphics[scale=0.7]{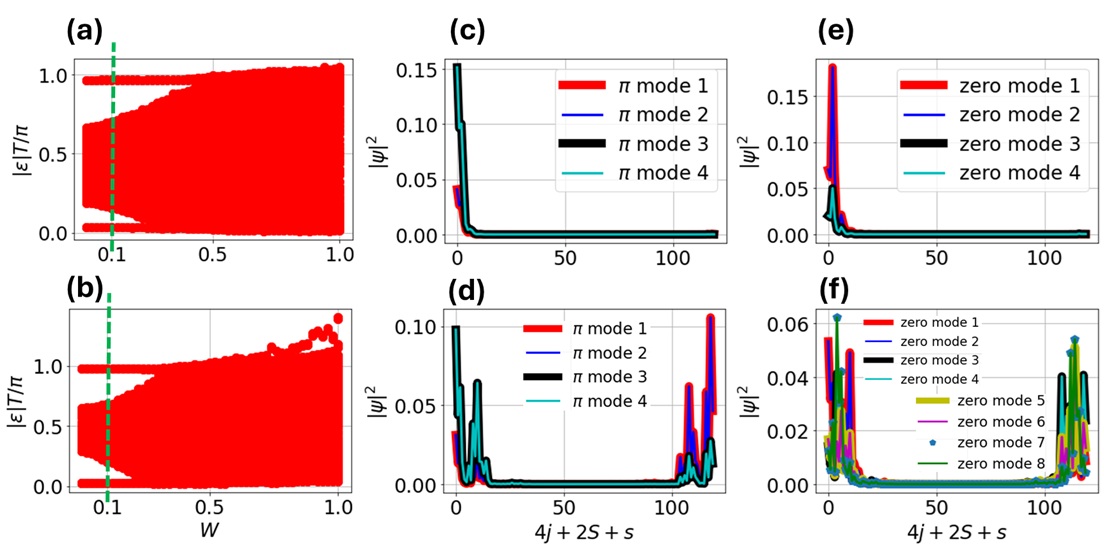}
        \caption{(a,b) The disorder averaged quasienergy spectrum at a varying disorder strength $W_P\equiv W$ for (a) model 1 and (b) model 2. The vertical dashed line marks the value of $W$ used in panels (c) to (f). (c,d) The corresponding spatial profiles of all $\pi$ modes at $W=0.1$ for (c) model 1 and (d) model 2. (e,f) The corresponding spatial profiles of all zero modes at $W=0.1$ for (e) model 1 and (f) model 2. The remaining parameter values are taken as $N=30$, (a,c,e) $\overline{\Delta}_0=0.5$, $\overline{\delta}_0=1$, $J_0=\sqrt{1.8^2+\Delta_0^2}$, $j_0=\sqrt{2.5^2+\delta_0^2}$, $\overline{\omega}=0.05$, (b,d,f) $m=1.2\pi$, $\overline{J}_a=\overline{J}_b=-\frac{\overline{j}_a}{m}=\frac{\overline{j}_b}{m}= \sqrt{1^2 + 0.6^2}$, $\overline{\Delta}_a=\overline{\Delta}_b=-\frac{\overline{\delta}_a}{m}=\frac{\overline{\delta}_b}{m}= 0.6$, $\tilde{\omega}=0.01\pi$. Each data point is averaged over $20$ disorder realizations.}
        \label{fig:dis1}
    \end{figure*}
\end{center}

To incorporate disorder, we make each parameter site-dependent according to $P\rightarrow P_j$, where $P\in \left\lbrace J_a, J_b,\Delta_a, \Delta_b, j_a, j_b, \delta_a, \delta_b \right\rbrace$, $P_j$ is randomly drawn from the interval $[\overline{P}-W_P \overline{P},\overline{P}+W_P \overline{P}]$, and $W_P$ is the disorder strength for parameter $P$. Our results are summarized in Fig.~\ref{fig:dis1}, where the same disorder strength $W_P=W$ is taken for all system parameters. In particular, Fig.~\ref{fig:dis1}(a,b) demonstrates that the quasienergy gaps remain open around $|\varepsilon|=0$ and $|\varepsilon|=\pi/T$, thereby preserving all topological edge modes, for up to $20\%$ in disorder. Meanwhile, Fig.~\ref{fig:dis1}(c) to (f) depict the disorder averaged spatial profiles (Eq.~(\ref{sprof})) corresponding to either the topological $\pi$ or zero modes, obtained by filtering in quasienergy excitation eigenstates closest to $\pi/T$ or zero respectively. 

\begin{center}
    \begin{figure}
        \centering
        \includegraphics[scale=0.45]{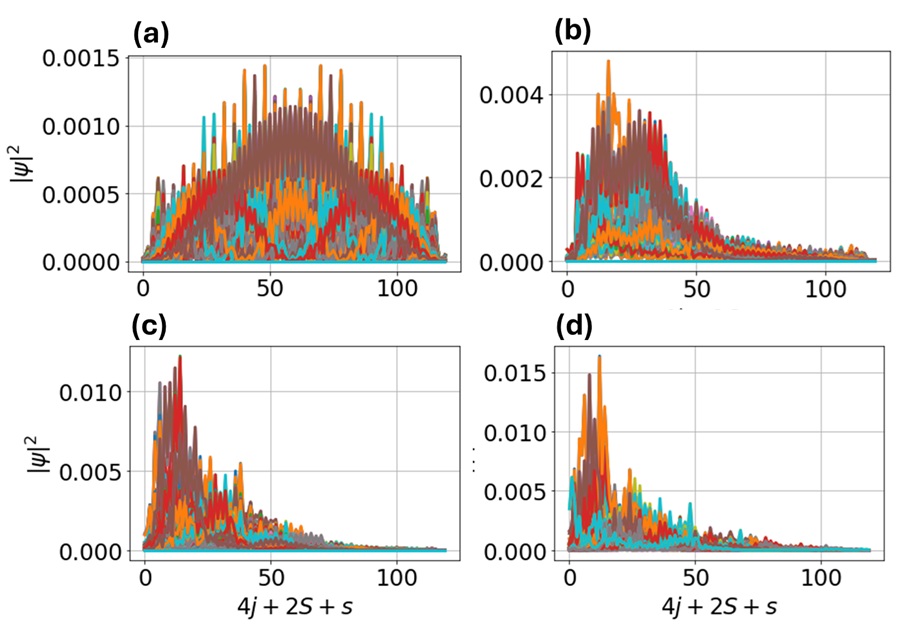}
        \caption{The disorder averaged spatial profiles of all nontopological bulk modes for model 1 at (a) $W=0$, (b) $W=0.1$, (c) $W=0.2$, (d) $W=0.4$. All other parameters are the same as in Fig.~\ref{fig:dis1}(a).}
        \label{fig:dis2}
    \end{figure}
\end{center}

Despite the similar robustness of all topological edge modes against disorder, both models display a strikingly different edge mode behavior. That is, while the topological modes in model 2 could be localized at either the left or right end, all topological modes in model 1 are localized at the left end. This feature suggests the disorder-induced revival of the NHSE in model 1. To further confirm this point, we plot the spatial profiles of all nontopological bulk modes only in Fig.~\ref{fig:dis2} for increasing disorder strength. Indeed, in the absence of disorder (panel (a)), all nontopological bulk modes are delocalized due to nonzero onsite bosonic frequency. Interestingly, when disorder is present, these nontopological bulk modes become more localized to the system's left end as the disorder strength increases. By contrast, for model 2, increasing the disorder strength has no qualitative impact on the spatial profiles of the nontopological bulk modes (not shown). That is, under the parameter values considered in Fig.~\ref{fig:dis1}(b,d,f), all nontopological bulk modes remain delocalized regardless of the disorder strength, suggesting the absence of disorder-induced revival of the NHSE. 

\section{Concluding remarks}
\label{conc}

In this paper, we present a class of periodically driven modified BKC that incorporates two sites per unit cell. Our driving scheme involves periodically switching the system Hamiltonian between two alternating forms at every half driving period. As case studies, two specific models are considered by imposing different constraints on the system parameter values. The first model represents a minimal topologically nontrivial Floquet modified BKC that still supports both topological zero and $\pi$ modes, whereas the second model is slightly more intricate but has the ability to support a controllable number of zero and $\pi$ modes. Moreover, all topological edge modes are found to demonstrate considerable robustness against spatial disorder.

While both models are described by Hermitian many-body Hamiltonians, the corresponding excitation Hamiltonians are generally nonHermitian. As the presence of topological zero and $\pi$ modes depends on the topology of the one-period time evolution of the corresponding excitation Hamiltonian, its studies involve the spectral analysis of some nonHermitian matrix. Consequently, physical features with no static counterparts such as the exotic NHSE, corresponding to all eigenstates being localized at a system end, emerge in both models. In particular, NHSE is present for both models in the absence of onsite bosonic frequency. The presence of onsite bosonic frequency completely destroys NHSE in model 1, but the presence of spatial disorder gradually revives it. Meanwhile, for model 2, NHSE may or may not survive in the presence of onsite bosonic frequency depending on the parameter values considered, but it is not qualitatively impacted by the presence of spatial disorder.

Our findings are expected to generate interest in studies of Floquet BKC and its variations. In particular, the two models presented in this work may serve as promising bases for more sophisticated models. For example, extending both models to a two-dimensional (2D) lattice may yield richer nonHermitian models that support a second-order topological phase and/or a first-order topological phase under a different symmetry class. Alternatively, the addition of terms with higher-powers in bosonic operators may turn the system effectively interacting, thereby allowing one to explore the interplay between interaction effect and nonHermitian physics. Apart from these potential theoretical future studies, Floquet BKC and its variations are also relevant to experimentalists as they serve as promising means to effectively realize Floquet nonHermitian topological phases from an otherwise Hermitian system. Finally, the topological zero and $\pi$ modes obtained in our models, combined with some parallel between BKC and its fermionic counterpart, may motivate explorations on their quantum computing prospects.

\end{document}